\documentstyle[11pt,paspconf,epsf]{article}

\begin{document}

\newcommand{\lx}{\mbox{$L_{\rm X}$}}
\newcommand{\lxlbol}{\mbox{$L_{\rm X}/L_{\rm bol}$}}
\newcommand{\ross}{\mbox{$N_{R}$}}
\newcommand{\vsini}{\mbox{$v \sin i$}}
\newcommand{\kms}{\mbox{km\,s$^{-1}$}}

\title{X-rays from Open Clusters}
\author{R. D. Jeffries}
\affil{Department of Physics, Keele University, Keele, Staffordshire
ST5 5BG, UK}

\begin{abstract}
The present state of X-ray observations of cool stars in 
coeval open clusters is reviewed.  
Concentrating on {\em ROSAT} results for solar-type stars, the available
observational dataset is summarized along with those details of the
evolution of X-ray activity of low mass stars which have been {\em
firmly} established as a result. Observational questions which are as
yet unresolved are then addressed, including the origin of
``supersaturation'' and whether observations of one cluster can
represent the X-ray properties of all clusters at the same age.  The
role of high spatial resolution X-ray imaging as a tool for identifying
cluster members is highlighted and the prospects for future
developments with AXAF and XMM are discussed.

\end{abstract}
           

\section{Introduction}
Open clusters are perhaps the best laboratories in which to study the
evolution of stellar properties. One might hope they contain
stars with a variety of masses, but with roughly the same
age, distance and composition. It is not surprising then that much X-ray
satellite time has been devoted to the study of open clusters in an
attempt to elucidate the roles that rotation, age, composition,
binarity, mass and initial conditions play in determining the levels of
X-ray emission from a star, if indeed such a deterministic approach is
possible.  

A major achievement of the {\em Einstein} observatory in the
1980s was to show that solar analogues in young clusters could exhibit
coronal X-ray activity orders of magnitude greater than our own Sun and
that this activity declined as the fast
rotating young stars were magnetically braked and 
lost their initial angular momentum - the {\em age-rotation-activity
paradigm} (ARAP).
A clear goal for the stellar X-ray astronomer would
be to describe the history of coronal activity in our own solar system,
but as we shall see, this is not yet possible in detail. Furthermore, a
comprehensive empirical understanding of X-ray luminosity may provide a
route to studying the age distributions of other stellar populations.
In the era of the {\em ROSAT} satellite the ARAP has been largely
confirmed, although various details remain puzzling. Confidence in the
ARAP means that X-ray observations can now be used as a tool to find and
study the low mass members of open clusters, in circumstances where optical
methods of membership selection are difficult or nearly useless.

\section{The ROSAT Era}

The {\em Einstein} observatory Imaging Proportional Counter (IPC)
showed that stars throughout the H-R diagram emit X-rays, particularly
those in close binary systems (RS~CVns) and the low mass stars of young
open clusters like the Pleiades and Hyades.
X-ray luminosity in cool stars with convective
envelopes clearly declined with age (Micela et al. 1990), but it had
become clear that the primary determinant of X-ray activity was
rotation rate.  This was interpreted as a natural consequence of the
dynamo mechanism whereby the magnetic fields which confine and heat
coronae, are amplified by rotation and convection. Indeed, X-ray
activity, expressed as the ratio of \lx\ to bolometric luminosity
(\lxlbol), was found to be even better correlated with Rossby number
(\ross), the ratio of rotation period to convective turnover time
(Dobson \& Radick 1989). The decline of X-ray activity with age was
then simply explained in terms of the angular momentum loss (AML) suffered
by young, single stars, with tidally locked, short period binary stars
remaining a high \lx\ polluting factor due to their continued fast
rotation.  
The discovery of spreads in
rotation rate among low mass stars at the same age (Stauffer 1991)
could then also explain why the X-ray luminosity functions (XLFs) of
the Pleiades and Hyades showed some overlap.  X-ray emission from
higher mass stars was either interpreted as due to a shallow convective
layer (early F stars), a lower mass binary companion (A stars) or
intrinsic emission from a shocked, radiatively driven wind (early B
stars).

The {\em Einstein} observations of open clusters were only partially
satisfactory. The relatively low sensitivity meant that many cluster
members (especially K and M stars) were undetected and much of the
analysis relied on statistical treatments of upper limits, with
consequently uncertain XLFs. Many questions were left hanging, such as:
If the ARAP operates in clusters, why is there such a small range of
\lx\ in the Pleiades where there is more than an order of magnitude
spread in rotation rate? What happens to the X-ray activity of stars
younger than the Pleiades ($\sim 100$\,Myr) and older than the Hyades
($\sim 700$\,Myr)? How much of the spread in X-ray activity at a given
age can be attributed to short timescale variability or magnetic
activity cycles? Is the activity of one cluster necessarily
representative of all clusters at the same age?  The launch of {\em
ROSAT} in 1990 offered the opportunity to answer these questions. Its
Position Sensitive Proportional Counter (PSPC) and High Resolution
Imager (HRI) had greater sensitivity, higher spatial resolution and in
the case of the PSPC, more spectral resolution than the IPC.

Table~1 is an update from the reviews of Caillault (1996) and Randich
(1997), which summarizes the {\em ROSAT} dataset on clusters (older than
10\,Myr). References to published results are given or if unpublished,
the PI on the observation is indicated. Ages and distances are adopted
from the Lyng\aa\ (1987) catalogue and should be treated with caution!
There are now deep, reasonably consistently calibrated soft X-ray
observations of more than 25 open clusters. This massive database has
answered most of the questions posed by {\em Einstein}, but yielded a
number of new mysteries. In particular the XLFs of F G and K stars are
now well determined, with few upper limits, in several open clusters at
ages from 30\,Myr to 600\,Myr.
 
\begin{center}
\begin{table}
\scriptsize
\caption{Open clusters observed with {\em ROSAT} (ages and distances
from Lyng\aa\ 1987).}
\begin{tabular}{lccll}
\hline Cluster & Log Age & Distance & {\em ROSAT} & References \\ &
(yr)& (pc) & Instrument\tablenotemark{a} & \\ \hline IC 2602 & 7.00 &
155 & PSPC (R) & Randich et al. 1995, A\&A, 300, 134\\ NGC 2232 & 7.35
& 400 & HRI (P) & PI Prosser \\ Col 140 & 7.35 & 300 & HRI (P) & PI
Prosser; PI Theissen\\ IC 2391 & 7.56 & 140 & PSPC (P) & Patten \&
Simon 1993, ApJ, 415, L123 + \\ & & & & Patten \& Simon 1996, ApJS,
106, 489\\ & & & HRI (P) & Simon \& Patten 1998, PASP, 501, 624\\ IC
4665 & 7.56 & 430 & HRI (P) & Giampapa et al.  1998, ApJ, 501, 624\\ &
& & HRI (P) & PI Giampapa\\ NGC 2451 & 7.56 & 220 & HRI (P) & PI
Schmitt; PI Huensch\\ Blanco 1 & 7.70 & 190 & HRI (P) & Micela et
al. 1999, A\&A in press \\ Alpha Per & 7.71 & 170 & PSPC (R) & Randich
et al. 1996, A\&A, 305, 785 \\ & & & PSPC (P) & Prosser et al. 1996,
AJ, 112, 1570\\ NGC 2547 & 7.76 & 400 & HRI (P) & Jeffries \& Tolley
1998, MNRAS, 300, 331\\ NGC 2422 & 7.89 & 480 & PSPC/HRI & Barbera et
al. 1996, CSSS9, p.355\\ Pleiades & 7.89 & 125 & PSPC (P) & Stauffer et
al. 1994, ApJS, 91, 625 + \\ & & & & Gagn\'{e} et al. 1995, ApJ, 450,
217\\ & & & PSPC (P) & Micela et al. 1996, ApJS, 102, 75\\ & & & PSPC
(S) & Schmitt et al. 1993, A\&A, 277, 114\\ & & & HRI (P) & Harnden et
al. 1996, CSSS9, p.359\\ & & & HRI (P) & PI Prosser \\

Stock 2 & 8.00 & 320 & HRI (P) & PI Sciortino\\ NGC 2516 & 8.03 & 440 &
PSPC (P) & Dachs \& Hummel 1996, A\&A, 312, 818 \\ & & & PSPC (P) &
Jeffries et al. 1997, MNRAS, 287, 350 \\ & & & HRI (P) & PI Micela\\
NGC 1039 & 8.29 & 440 & HRI (P) & PI Simon\\ NGC 6475 & 8.35 & 240 &
PSPC (P) & Prosser et al. 1995, AJ, 110, 1229\\ & & & PSPC (P) & James
\& Jeffries 1997, MNRAS, 292, 252\\ & & & HRI (P) & PI James\\ NGC 7092
& 8.43 & 270 & HRI (P) & PI Favata; PI Micela\\ NGC 3532 & 8.54 & 500 &
HRI (P) & PI Simon\\ Coma Ber & 8.60 & 86 & PSPC (P) & Randich et
al. 1996, A\&A, 313, 815 \\ IC 4756 & 8.76 & 400 & HRI (P) & Randich et
al. 1998, A\&A, 337, 372\\ NGC 6633 & 8.82 & 320 & PSPC (P) & Totten et
al. these proceedings\\ Hyades & 8.82 & 48 & PSPC (S) & Stern et
al. 1992, ApJ, 399, L159; \\ & & & & Stern et al. 1995, ApJ, 448, 683
\\ & & & PSPC (P) & Pye et al. 1994, MNRAS, 266, 798\\ & & & PSPC (P) &
Stern et al. 1994, ApJ, 427, 808\\ & & & PSPC (P) & Reid et al. 1995,
MNRAS, 272, 828\\ & & & HRI (P) & PI Walter\\ Praesepe & 8.82 & 180 &
PSPC (R) & Randich \& Schmitt 1995, A\&A, 298, 115 + \\ & & & & Barrado
et al. 1998, ApJ in press\\ NGC 6940 & 9.04 & 800 & PSPC (P) & Belloni
\& Tagliaferri 1997, A\&A, 326, 608\\ NGC 752 & 9.04 & 400 & PSPC (P) &
Belloni \& Verbunt 1996, A\&A, 305, 806\\ NGC 3680 & 9.26 & 800 & HRI
(P) & PI Tagliaferri \\ IC 4651 & 9.38 & 710 & PSPC (P) & Belloni \&
Tagliaferri 1998, A\&A, 335, 517\\ M67 & 9.60 & 720 & PSPC (P) &
Belloni et al. 1993, A\&A, 269, 175\\ NGC 188 & 9.70 & 1550 & PSPC (P)
& Belloni 1997, MemSAIt, 68, 993\\ \hline
\end{tabular}
\tablenotetext{a}{P - pointed observation, R - raster scan observation,
S - all-sky survey observation}

\end{table}
\end{center}

\subsection{The Age-Rotation-Activity Paradigm}

\begin{figure}[ht]
\plotone{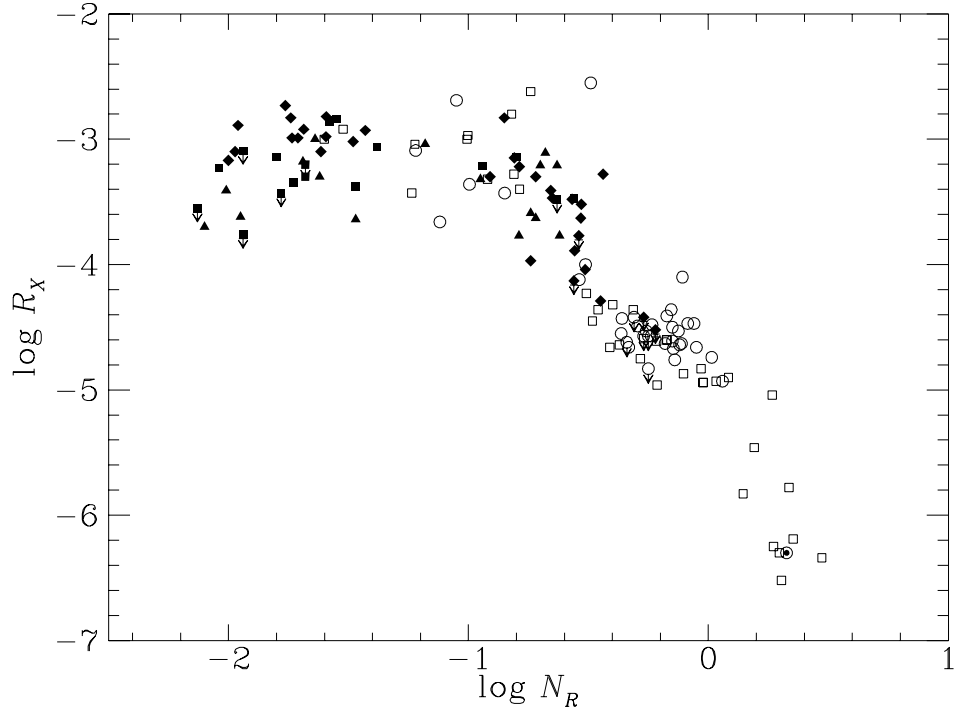}
\caption{X-ray activity (\lxlbol) as a function of \ross\ for late type
stars (F5-M5). Data are for IC2391 (filled triangles), $\alpha$ Per
(filled squares), Pleiades (filled diamonds), Hyades (open circles) and
field stars (open squares). (From Patten \& Simon 1996)}
\end{figure}

At the same time as the emergence of new {\em ROSAT} results,
observations of rotational broadening (\vsini -- see Stauffer 1991) 
or photometrically determined rotation periods
({\em e.g.} Prosser et al. 1995 and refs. therein) have allowed the
evolution of stellar rotation with age to be studied in detail.  The
prevailing interpretation of these data is that young PMS stars
contract and spin-up as they approach the ZAMS after magnetically
uncoupling from any circumstellar disk. A variety of disk coupling
lifetimes leads to more than an order of magnitude spread in rotation
rates at the ZAMS (Bouvier et al. 1997).  Meanwhile, AM is lost via a
magnetized stellar wind and in order to maintain a big spread in
rotation rates, saturation of the AML rate is needed in the fastest
rotators.  Once on the ZAMS, with a constant moment of inertia, stars
lose AM at a mass-dependent rate. Spindown timescales vary from a few
tens of Myr for G stars to a few hundred Myr and longer for K and M
stars. To reproduce this in models requires that the rotation rate
above which AML saturation takes place is such that saturation occurs
at {\em roughly} constant \ross\ (Krishnamurthi et al. 1997).

Coronal activity depends on rotation through the dynamo process, hence
a correlation between X-ray activity (usually expressed as a fraction
of bolometric luminosity - \lxlbol) and rotation is expected. This was
well demonstrated by Stauffer et al. (1994) in the Pleiades, Randich et
al. (1996a) in the $\alpha$ Per cluster and Stauffer et al. (1997) in IC
2391/2602. If one looks at G stars in the Pleiades for instance, stars
with \vsini$<15$\kms\ show an order of magnitude spread in \lxlbol,
with the fastest rotators having the highest activity. This is even more
clearly seen when using the more precise \vsini\ measurements of Queloz
et al. (1998). However, the surprising result was that above a \vsini\
of $\sim15$\kms, \lxlbol\ seems to reach a saturated plateau value of
$10^{-3}$. The reason for this saturation is still unknown. It may be
caused by filling of the available coronal volume with plasma or it may
be due to a negative feedback in the dynamo mechanism itself.
Saturation explains why there is a limited range in \lx\ among Pleiads
of a given mass, even though rotation rates vary by more than a factor
10.

\begin{figure}
\plotone{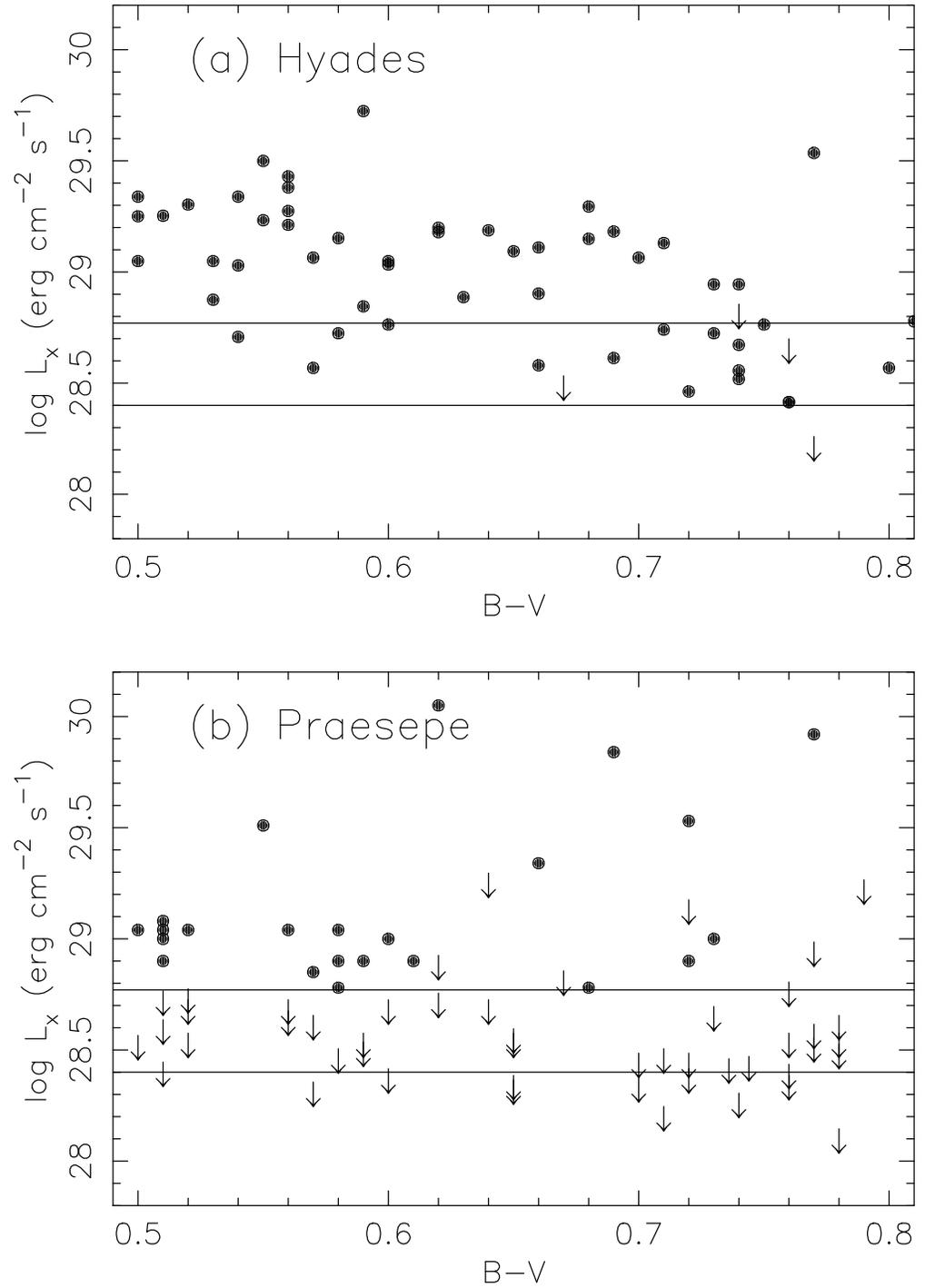}
\caption{The \lx\ distribution for solar-type stars in the (a) Hyades
and (b) Praesepe. The
two horizontal lines indicate the approximate sensitivity limits of the
{\em ROSAT} surveys for the Hyades (lower line) and Praesepe (upper line)
The average level of X-ray emission in Praesepe is significantly lower
than the Hyades (adapted from Barrado et al. 1998).}
\end{figure}

There is a considerable scatter around the rotation-activity
correlation when stars with higher or lower masses are added.
Following earlier work on magnetic activity by 
Dobson \& Radick (1989), it has been found that including a parameter
describing the convective cell turnover time, significantly improves
correlations and unifies the view of dynamo generated coronal
activity. The new parameter of choice is \ross. Fig.1 (from Patten \&
Simon 1996) shows data from IC 2391, $\alpha$ Per, the Pleiades, Hyades
and main sequence field stars. There is still some scatter around this
relation, which may be due to time variability/activity cycles, but
there appears to be a flattening of activity at $\log \ross
=-0.8\pm0.2$, {\em irrespective of spectral type}, which compares
favourably with similar analyses by Stauffer et al. (1997) and Queloz
et al. (1998). Because convection zones get deeper and convective
turnover times longer in lower mass stars, a uniform saturation \ross\
indicates a rotation period for saturation that gets longer at lower
masses. It is tempting to speculate (see Krishnamurthi et al. 1998)
that the coronal saturation coincides with the mass dependent
saturation of AML that seems to be required in rotational evolution
models (Barnes \& Sofia 1996, Krishnamurthi et al. 1997). As yet, the
details of (for instance) internal differential rotation are too
uncertain for this speculation to be confirmed.

The age dependence of X-ray emission is thought to arise solely as a
consequence of the rotation dependence. A clear demonstration of the
effect is given by Caillault (1996 - Fig. 1). The XLFs of young
clusters have a spread {\em approximately consistent} (see below) with
the rotation-activity relation (and saturation) and have peak and
median values of \lx\ that decrease with age at a mass-dependent rate.
A telling confirmation of the ARAP is that the spectral type of stars
at which the saturation value of \lxlbol\ is achieved gets cooler in
older clusters as the the higher mass stars spindown below their
saturation threshold. Unfortunately, the convergence of rotation rates
in solar-type stars by the time they reach the Hyades age, means that
at the moment we cannot say much about the history of our own Sun's
activity prior to this epoch. However, the dispersion in rotation and
hence coronal emission remains for somewhat longer in lower mass stars
(Stern et al. 1995; see also Hawley in these proceedings), because of
their longer spindown timescales.

\subsection{Time variability}

Prior to {\em ROSAT} measurements little was known about coronal
variability on timescales of months or years and whether it might be
responsible for the spreads in \lx\ seen in young clusters.  Coronal
variability is reviewed in these proceedings by Stern, so only a brief
summary is given here. Solar coronal variability is $>$ a factor of 10
in soft X-rays during its activity cycle (Peres et al.-- these
proceedings), but how do younger stars behave?  The Hyades and Pleiades
now have multiple epoch X-ray observations. Both Gagn\'{e} et
al. (1995) and Micela et al. (1996) show that variability in Pleiades
stars is smaller than solar variations on timescales of 6 months to ten
years. Perhaps 20\% to 40\% are variable by as much as a factor 2. This
might be thought due to the ceiling on X-ray emission
provided by saturation, but Stern et al. (1995) show that similar
results hold for G and K stars in the Hyades, which are not rotating
fast enough for saturation. Stern et al. suggest that the difference
in the behaviour of the Sun and younger stars might result from the
action of a ``turbulent'' dynamo in their convection zones, which does
not exhibit the cyclical behaviour seen in the Sun.  In any case it
could be considered fortunate that time variability is not sufficient
to cause the spreads in activity seen in young clusters and cannot
disguise the ARAP. Unfortunately the \lx\ of single solar-type stars in
older clusters is too low for {\em ROSAT} -- so we still do not know at
what age solar-type variability sets in, although there are numerous
examples of short-period binary systems which have been detected at the
expected levels of emission ({\em e.g.} Belloni 1997 and
refs. therein).

\section{Unsolved Mysteries}

Despite the success of the ARAP in describing most X-ray observations
of open clusters, there remain a number of problems that either require
a rethink or extension of the paradigm. Among these are the possibility
of a ``third parameter problem'', the phenomenon of ``supersaturation''
and whether one cluster at a given age has X-ray properties
representative of all similar clusters.

That rotation and spectral-type (or mass) alone might be insufficient
to determine X-ray activity was postulated by Micela et
al. (1996). They claim that the spread in X-ray activity among slowly
rotating G/K Pleiades stars is too large to be accounted for by
uncertainties in flux, inclination angle or variability. Similarly,
Fig.10 in Stern et al. (1995) shows more than an order of magnitude
spread in \lx\ for F8-G5 Hyades stars, even though their rotation rates
are thought to be almost uniform and their variability is demonstrated
to be small in the same paper. Micela et al. suggest that the internal
rotation profile may be the ingredient that is missing from the ARAP.
I believe that this problem may yet be due to a combination of errors
in \vsini\ measurements (in the case of K stars) and grouping stars
with a range of convective turnover times (for late F and G
stars). Using more accurate \vsini\ measurements, Queloz et al. (1998)
show that \lxlbol\ is well correlated with \ross$/\sin i$, with only
about a factor of 3 spread, which is perfectly consistent with
uncertain inclination angles, X-ray flux variability and errors. The
key to resolving this issue definitively is to obtain accurate rotation
periods for many more stars, rather than new X-ray observations.

``Supersaturation'' is the observed phenomenon that at very fast
rotation rates, \lxlbol\ appears to decrease by a factor of 3-5 below
the canonical saturation limit of $10^{-3}$ (Prosser et al. 1996;
Randich 1998).  Because it affects only the fastest rotating late-type
stars (\vsini$>100$\kms) 
and there are only a few ($\sim 10$) of these in the very young
$\alpha$ Per and IC 2391/2602 clusters, it is still not clear whether
supersaturation sets in at a particular rotation rate or a
particular \ross.  The latter is not favoured by observations of dMe
stars (see James et al. in these proceedings).  An explanation for
supersaturation is not obvious. Randich (1998) suggests that it may
represent a fall-off of the dynamo mechanism itself or perhaps a shift
in the distribution of the radiative losses out of the {\em ROSAT} band
to either hotter or cooler temperatures. The latter explanation may
draw some support from the few X-ray spectra available for stars in the
Pleiades (Gagn\'{e} et al. 1995) which indicate that the fastest
rotating G stars have hotter coronae than the slow
rotators. Unfortunately there are almost no nearby field stars that
rotate fast enough to exhibit supersaturation {\em and} have accurate
X-ray spectra that could test this hypothesis.  Perhaps a more likely
explanation is centrifugal shrinkage of the available coronal volume at
very high rotation rates.

Most of the early interpretation of X-ray observations relied on the
assumption that one cluster was representative of all clusters at the
same age. In the {\em ROSAT} era, this assumption can be tested (see
Table~1). One of the most important results of this decade, illustrated
in Fig.3, was that the coronal activity of low mass stars (especially
solar-type) in Praesepe was significantly lower on average than those in
the Hyades at the same age, although peak \lx\ values were similar
(Randich \& Schmitt 1995). Explanations range from contamination by
non-members in the Praesepe sample, differing binary fractions or
orbital distributions, differing initial conditions (specifically the
AM distribution) or rotational evolution (perhaps influenced by
composition differences -- Jeffries et al. 1997). Mermilliod (1997)
shows that the rotation rate distributions in the two clusters are
similar and Barrado y Navascu\'{e}s et al. (1998) find that
contamination with non-members is unlikely.  Although much more work
needs to be done on identifying and parameterizing binaries in Praesepe,
a genuine explanation remains elusive.  Observationally, the situation
has now become more confusing. Randich et al. (1996b) find that the
pattern of X-ray emission in the F/G stars of the similarly aged Coma Berenices
cluster resembles the Hyades rather than Praesepe, whereas the X-ray
activity of F/G stars in NGC 6633 and F stars in IC 4756 is less than
the Hyades (Totten et al. these proceedings, Randich et al. 1998).  Deeper
observations of Praesepe, NGC 6633 and IC4756 to remove the upper
limits will certainly assist interpretation, as will careful optical
work to identify spectroscopic binaries. AXAF offers the opportunity to
extend these tests to more distant open clusters with a range of ages,
composition and richness.

\section{High Spatial Resolution}

\begin{figure}[h]
\plotone{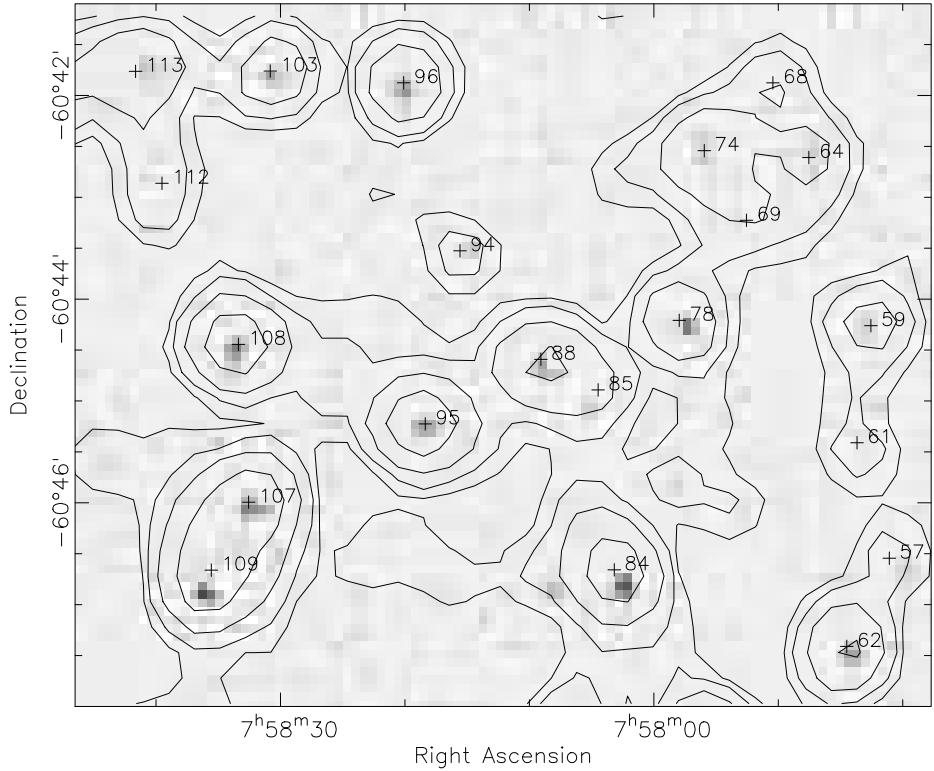}
\caption{The core of NGC 2516 as seen by the PSPC (contours) and by the
HRI (greyscale). The numbered sources are from Jeffries et al. 1997.}
\end{figure}

\begin{figure}[h]
\plotone{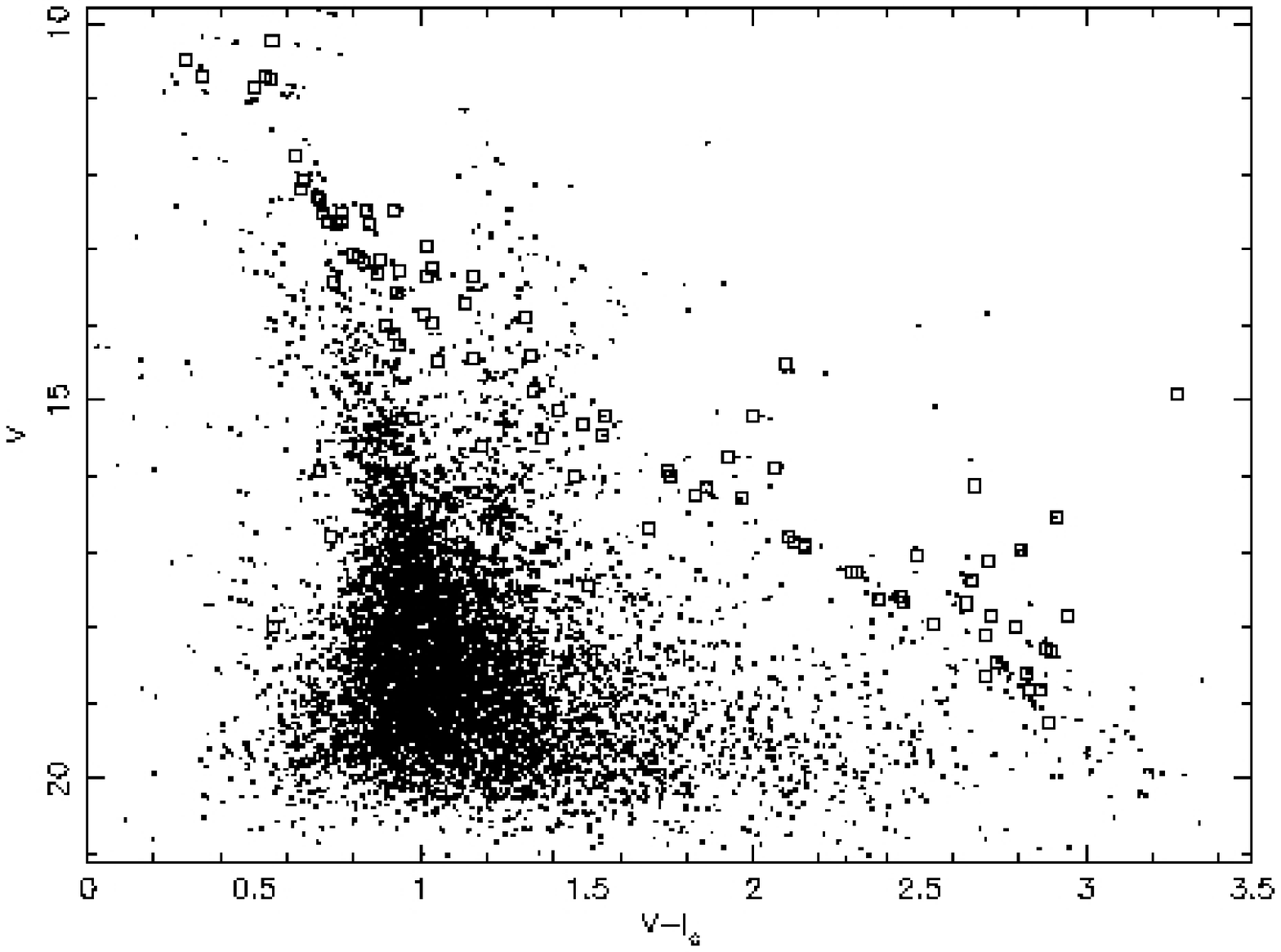}
\caption{A colour-magnitude diagram for NGC 2547 together with those
stars residing inside HRI X-ray error circles (squares).}
\end{figure}

High spatial spatial resolution is useful in several distinct ways in
studies of open clusters. First, as one moves towards more distant open
clusters, the angular separation between cluster X-ray sources becomes
small and accurate analysis becomes difficult as PSFs start to
overlap. A good example is NGC 2516 (Jeffries et al. 1997), where 159
X-ray sources were found within 20$^{'}$ of the centre of a PSPC
pointing, and multiple PSF fitting (analogous to DAOPHOT for optical
photometry) was required. This cluster has been re-observed at about 8
times the spatial resolution (but less sensitivity) with the HRI
(FWHM$\sim3^{''}$) and the results are shown in Fig.3.  Essentially all
the sources found by the HRI are also extracted from the PSPC data at
the right positions. At this sensitivity NGC 2516 represents the most
distant cluster (for its age) which could be studied with the spatial
resolution of the PSPC. As nowhere near all the cluster members were
detected by the PSPC one would like to go deeper, but it is clear that
HRI-like spatial resolution will be required to do this. The problem
will be more acute for more distant clusters with a similar
``richness''.

A different problem is the identification of X-ray sources with their
optical counterparts. In clusters close to the Galactic plane (and more
distant clusters will tend to be so), there may be several candidate
stars within each X-ray error circle. Clearly, improvements in spatial
resolution concomitantly reduce the number of possible optical
counterparts. For most nearby clusters the increase in resolution from
PSPC to HRI is not too important from this perspective ({\em e.g.}
Simon \& Patten 1998), but for others near the Galactic plane or for
deeper studies with fainter possible optical counterparts, higher
spatial resolution than offered by the PSPC is absolutely essential
({\em e.g.} Giampapa et al. 1998).

A related problem that plagues studies (not just in X-rays) of low mass
stars in open clusters is a lack of firm membership lists for faint
stars. Proper motions and photometric selection can be useful among
nearby clusters, but as one moves to more distant clusters, which tend
to be projected against the Galactic plane, both background
contamination and small proper motions become problematic. An
alternative approach, which relies on the ARAP, is to find low mass
members by X-ray selection, because the contrast in X-ray flux between
young, low mass cluster stars and a general field population is
large. This approach has now been used in a number of young clusters
where optically selected membership catalogues are either absent or
very uncertain ({\em e.g.} Prosser et al. 1994; Patten \& Simon
1996). A combination of photometric selection plus small X-ray error
circles can be especially powerful and if the X-ray survey is deep
enough then one can be reasonably sure that the selected members are a
complete sample, rather than biased towards high activity levels.
A recent example is shown in Fig.4. NGC 2547 is a young cluster
observed with the HRI (Jeffries \& Tolley 1998). The $\sim 6$ arcsec
error circles contain one star on average and the majority of these
form a sequence along the expected locus of the cluster in a
colour-magnitude diagram. Less than one correlation is expected by
chance in this part of the CMD, so these stars must be genuine cluster
members. Furthermore, the X-ray survey is sufficiently deep that
the X-ray selected G and K stars are expected to be a complete sample
from which other investigations can be based.

\section{AXAF and XMM}

High resolution spatial imaging, moderate spectral resolution and
increased effective area over {\em ROSAT} are the key attributes of the
AXAF satellite. The ACIS instrument is capable of 1 arcsec resolution
with spectral resolution of $\sim 20$ at $\sim 2$\,keV over a 16
arcmin field.  The ACIS sensitivity threshold will be a factor
of 3-5 better than the PSPC/HRI for the same exposure time.  The HRC
instrument has even better spatial resolution over a larger 31 arcmin
field but little spectral resolution. XMM has a larger
overall effective area and the capability of providing spectral
resolution of several hundred for individual targets. Together, the
capabilities of the two instruments will make significant
advances in our understanding of the X-ray evolution of cool stars in
open clusters.
\begin{itemize}

\item For nearby clusters (the Pleiades and NGC 2516 in AO-1),
X-ray emission can be probed further down the main sequence to see
if the nature of the dynamo changes in fully convective stars or even
brown dwarfs.

\item The spectral resolution will allow detailed modelling of the
X-ray spectra. At present, a single, crude spectral model is normally assumed
to convert from count rate to flux. We will be able to see if coronal
temperatures continue to increase in the most rapidly rotating stars
and whether a shift of flux to higher energies causes supersaturation
in the {\em ROSAT} pass band. XMM will be able to provide detailed, high resolution
spectroscopic studies of even low activity stars in both the Hyades and Pleiades
at \lx\ thresholds of roughly $10^{28}$ and $10^{29}$\,erg\,s$^{-1}$
respectively, in reasonable exposure times.

\item For the first time, the detection of 
X-ray emission from main sequence stars in clusters older than the
Hyades will be possible. The AXAF mission time should be long enough to
also get second epoch observations to study variability in these older
clusters.

\item AXAF should be able to study clusters out to distances of 1\,kpc
and the high spatial resolution means that even clusters with low $b$
will be accessible. Deeper observations of clusters at the same age as
the Hyades are required to remove the remaining ambiguities posed by
{\em ROSAT} upper limits. Several clusters at the same age as the
Pleiades ($\sim 100$\,Myr) and NGC 6475 ($\sim 300$\,Myr) can be
studied.

\item More distant, compact open clusters can be examined to find more
of the rare open clusters with ages between 10 and 40\,Myr, which are
so important in understanding the history of AML and circumstellar disks.

\end{itemize}

\acknowledgments

I would like to thank Giusi Micela, David Barrado y Navascu\'{e}s and Brian Patten for
providing data and figures for this review.

\end{document}